\documentclass{appolb}
\usepackage{epsfig}

\usepackage{subfigure}
\usepackage{epstopdf}
\usepackage{hyperref} 

\begin{document}
\title{Towards the assignments for $1^{1}D_{2}$ and $1^{3}D_{2}$ meson nonets
}
\author{Xue-Chao Feng $^{1}$\thanks{fxchao@zzuli.edu.cn/fxchao66@163.com}, Ke-Wei Wei $^{2}$\thanks{weikw@ihep.ac.cn}, Jie Wu $^{1}$\thanks{blueswj@163.com}, Xue-Zhen Zhai $^{1}$\thanks{wanzifsh@163.com}, Shi-Zhuo Wang $^{1}$\thanks{wsz@zzuli.edu.cn}
\address{$^{1}$ Department of Technology and Physics, Zhengzhou University of Light Industry, 450002 Zhengzhou, China
\\ $^{2}$ School of Science, Henan University of Engineering, 451191 Zhengzhou, China } }

\maketitle
\begin{abstract}
In this work, we investigate the mass spectrum of $1^{1}D_{2}$ and $1^{3}D_{2}$ meson nonets in the framework of the meson mass matrix and Regge phenomenology. The results are compared with the values from different phenomenological models and may be useful for the assignment of the $1^{1}D_{2}$ and $1^{3}D_{2}$ meson nonets in the future.

\end{abstract}
\PACS{11.55.Jy, 12.40.Yx, 14.40.Be}

\section{Introduction}

Quantum Chromodynamics (QCD) was established half a century ago as a theory of strong interactions. In the past few years, QCD has achieved great success and continued to provide satisfactory explanations for many puzzling experimental results. However, we also need to face the fact objectively that there is still a long way to obtain a comprehensive understanding of strong interactions. One of the problems to be solved is that perturbative QCD is not applicable at low energy, and it is difficult to produce accurate results from non-perturbative calculations \cite{Blundell:1996as}. In this case, it is necessary to build models that can reveal the most important features of QCD, e.g., quenched lattice QCD \cite{Luscher:1996jn,Aoki:2009ji}, the Dyson-Schwinger formalism \cite{Maris:2003vk}, constituent quark models \cite{Gell-Mann:1964ewy}, and light cone QCD \cite{Lepage:1980fj,Buchalla:2008jp}. In addition, since mesons are an ideal laboratory for studying strong interactions in the strongly coupled non-perturbative regime, the study of the meson spectrum is of great significance for better understanding the dynamics of strong interactions \cite{Godfrey:1998pd,Li:2004gu,Koenigstein:2016tjw,Jafarzade:2022uqo}.

In this work, we will discuss the assignment of the $1^{1}D_{2}$ and $1^{3}D_{2}$ meson nonets. The masses and widths of the $1^{1}D_{2}$ and $1^{3}D_{2}$ meson nonets are listed in Table 1. Here, the values are taken from PDG \cite{ParticleDataGroup:2020ssz}.
\begin{center}
\indent\\ \footnotesize Table 1. The masses and widths of $1^{3}D_{2}$ and $1^{1}D_{2}$ meson nonets (in units of MeV). $\rho_{2}(???)$, $\omega_{2}(???)$, and $\phi_{2}(???)$ remain to be found in experiments.
\label{Tab:t1}
\begin{tabular}{llllllllll}
\\ \hline\hline
 State ($1^{1}D_{2}$)                 &mass               & width              & State ($1^{3}D_{2}$)              &mass            & width
\\
\hline
 $\pi_{2}(1670)$          &  $1670.6^{+2.9}_{-1.2}$  & $258^{+8}_{-9}$         &  $\rho_{2}(???)$   &
\\
 $\eta_{2}(1645)$         &  $1617\pm 5$             & $181\pm 11$             &  $\omega_{2}(???)$   &
\\
  $\eta_{2}(1870)$        &  $1842\pm 8$             & $225\pm 14$             &  $\phi_{2}(???)$   &
\\
$K_{2}(1770)^{\dag}$      &  $1773\pm 8$             & $186\pm 14$             &  $K_{2}(1820)^{\dag}$ &  $1819\pm 12$             & $264\pm 34$

\\ \hline
\end{tabular}
\end{center}

For the $1^{1}D_{2}$ meson nonet, the states $\pi_{2}(1670)$, $\eta_{2}(1645)$ and $\eta_{2}(1870)$ are well established. The state $\pi_{2}(1670)$ was observed in the reaction $\pi^{+}p\rightarrow p\pi^{+}\pi^{+}\pi^{-}$ half a century ago \cite{Bartsch:1968zz}. In 2002, the E852 Collaboration performed a partial-wave analysis of the reaction $\pi^{-}p\rightarrow \pi^{+}\pi^{-}\pi^{-}p$ and the $\pi_{2}(1670)$ was confirmed \cite{Chung:2002pu}. In 2005, three isovector $2^{-+}$ states $\pi_{2}(1670)$, $\pi_{2}(1880)$ and $\pi_{2}(1970)$ were seen in the $\omega \rho^{-}$ decay channel \cite{E852:2004rfa}. In Ref. \cite{COMPASS:2018uzl}, the state $\pi_{2}(1670)$ was measured with mass $1642^{+12}_{-1}$ MeV and width $311^{+12}_{-23}$ MeV. For the isoscalar, the Crystal Barrel Collaboration observed $\eta_{2}(1645)$, and $\eta_{2}(1870)$ in the reaction $p \bar{p}\rightarrow \eta \pi^{0}\pi^{0}\pi^{0}$ \cite{CrystalBarrel:1996bnu}. Subsequently, the $\eta_{2}(1645)$ and $\eta_{2}(1870)$ were studied by the WA102 Collaboration in the $a_{2}(1320)\pi$, $f_{2}(1270)\pi$, $a_{0}(1320)\pi$ channels \cite{WA102:1999ybu,WA102:1999lqn}. For the kaon sector, the assignment of $K_{2}(1770)$ remains interesting. The $K_{2}(1770)$ was first reported in the reaction $ K^{-}p \rightarrow pK^{-}\pi^{+}\pi^{-} $ and $ K^{-}p \rightarrow p\bar{K}^{0}\pi^{-}\pi^{0} $ \cite{Bartsch:1966zz}. Moreover, the evidence for two $J^{P}=2^{-}$ strange mesons was presented in the reaction $ K^{-}p \rightarrow K^{-}\pi^{+}\pi^{-}\pi^{0}p $ \cite{Aston:1993qc}. Unlike the $\pi_{2}(1670)$, $\eta_{2}(1645)$ and $\eta_{2}(1870)$, the kaon has non-diagonal C-parity, the spin-singlet ($1^{1}D_{2}$) and spin-triplet ($1^{3}D_{2}$) can mix to produce the physical states $K_{2}(1770)$ and $K_{2}(1820)$ \cite{Barnes:2002mu}. The cognition of the $1^{3}D_{2}$ meson nonet is even lower than that of the $1^{1}D_{2}$ meson nonet. Since $1^{3}D_{2}$ meson has the same spin angular momentum and orbital momentum as $1^{1}D_{2}$ and $1^{3}D_{1}$ meson, the mass of the corresponding states may have a large overlap. Moreover, the large decay widths also make it very difficult to find these states. Till now, only the kaon $K_{2}(1820)$ is listed in Table 1, the other members are still not established in the experiments. Therefore, both the theoretical and experimental analysis of the assignment for $1^{1}D_{2}$ and $1^{3}D_{2}$ meson nonets is required.

In this work, the mass spectrum of $1^{1}D_{2}$ and $1^{3}D_{2}$ meson nonets is investigated in the framework of the meson mass mixing matrix and Regge phenomenology. This work is organized as follows. The mass mixing matrix is described in section 2. The mass relations are derived from Regge phenomenology in section 3. The summary is given in section 4.

\section{Mass mixing matrix of isoscalar states}

In the quark model, mesons exist as $q\bar{q'}$ bound states of quark $q$ and antiquark $\bar{q'}$ (the flavors of $q$ and $\bar{q'}$ may be different). The $q\bar{q'}$ bound states can be classified based on the representations of SU(3) flavor group. In general, states with the same $J^{P}$ and additive quantum numbers can mix, so the bare isoscalar states can mix and form two physical states \cite{Li:2008mza,Li:2008et,Giacosa:2017pos}.

In the $n\bar{n}=(u\bar{u}+d\bar{d})/\sqrt{2}$ and $s\bar{s}$ basis, the mass matrix describing the mix of the physical isoscalar states can be expressed as
\begin{equation}
\label{Eq1}
M^2=\left(\begin{array}{cc}
M^2_{n\bar{n}}+2A_{nn} & \sqrt{2}A_{ns} \\
\sqrt{2}A_{ns}  & M^2_{s\bar{s}}+A_{ss}
\end{array}\right)
\end{equation}
with
\begin{equation}
\label{Eq2}
A_{nn}=\frac{\zeta}{m_{n}m_{n}},\quad A_{ns}=\frac{\zeta}{m_{n}m_{s}},\quad A_{ss}=\frac{\zeta}{m_{s}m_{s}},
\end{equation}
where $M_{n\bar{n}}$ and $M_{s\bar{s}}$ are the masses of states $n\bar{n}$ and $s\bar{s}$, respectively; $A_{nn}$, $A_{ns}$ and $A_{ss}$ are the mixing parameters, which describe the $q\bar{q}\leftrightarrow q'\bar{q'}$ transition amplitudes \cite{Brisudova:1997ag,Delbourgo:1998kg}; $\zeta$ is an SU(3)-invariant phenomenological parameter, $m_{n}$, $m_{s}$ are the masses constituent quarks ($n$ denotes the up or down quark).

For a given meson nonet, the physical isoscalar states $\Phi$ and $\Phi'$ are the eigenstates of the mass matrix $M^2$
\begin{equation}
\label{Eq3}
UM^2U^\dagger=\left(\begin{array}{cc}
M^2_{\Phi}&0\\
0&M^2_{\Phi'}
\end{array}\right),
\end{equation}
where $U$ is the unitary matrix, $M^2_{\Phi}$ and $M^2_{\Phi'}$ are masses of states $\Phi$ and $\Phi'$,  respectively.

From the realtions (\ref{Eq1}), (\ref{Eq2}) and (\ref{Eq3}), one has
\begin{equation}
\label{Eq4}
M^2_{n\bar{n}}+\frac{2\zeta}{m^{2}_{n}}+M^2_{s\bar{s}}+\frac{\zeta}{m^{2}_{s}}=M^2_{\Phi}+M^2_{\Phi'},
\end{equation}

\begin{equation}
\label{Eq5}
\left(M^2_{n\bar{n}}+\frac{2\zeta}{m^{2}_{n}}\right) \left(M^2_{s\bar{s}}+\frac{\xi}{m^{2}_{s}}\right)-\frac{2\xi}{m^{2}_{n}m^{2}_{s}}=M^2_{\Phi}
M^2_{\Phi'}.
\end{equation}

The relations (\ref{Eq4}) and (\ref{Eq5}) are the trace and determinant of relation (\ref{Eq3}), respectively.

Applying the relations (\ref{Eq4}) and (\ref{Eq5}) for the $1^{1}D_{2}$ and $1^{1}P_{1}$ meson nonets, we obtain the masses $M_{s\bar{s}({1^{1}D_{2}})}=1849.0\pm 8.1$ MeV or $1609.1\pm5.7$ MeV and $M_{s\bar{s}({1^{1}P_{1}})}=1423.9\pm8.7$ MeV or $1156.4\pm 8.1$ MeV. The masses $M_{s\bar{s}({1^{1}D_{2}})}=1609.1\pm5.7$ MeV and $M_{s\bar{s}({1^{1}P_{1}})}=1156.4\pm8.1$ MeV should be discarded since they are smaller than the masses of the isovector states $\pi_{2}(1670)$ and $b_{1}(1235)$. In this work, we use the quark masses $m_{n}=0.29$ GeV, $m_{s}=0.46$ GeV as extracted from meson spectroscopy \cite{Burakovsky:1998zk,Burakovsky:1998ct}.

Apart from the mixing of isoscalar states, for the kaons of $1^{1}D_{2}$ and $1^{3}D_{2}$, the spin-singlet and spin-triplet can mix to form physical states $K_{2}(1770)$ and $K_{2}(1820)$. The mixing can be expressed as

\begin{equation}
\label{Eq6}
K_{2}(1773)=K_{2}\left({1}^{1}D_{2}\right) \cos \theta_{K}+K_{2}\left({1}^{3} D_{2}\right) \sin \theta_{K},
\end{equation}
\begin{equation}
\label{Eq7}
K_{2}(1820)=-K_{2}\left({1}^{1}D_{2}\right) \sin \theta_{K}+K_{2}\left({1}^{3} D_{2}\right) \cos \theta_{K},
\end{equation}
where $\theta_{K}$ is the ($1^{1}D_{2}$)-($1^{3}D_{2}$) mixing angle. The same analysis process is used for axial-vector mesons \cite{Li:2006we}. The masses of $K_{2}(1^{1}D_{2})$ and $K_{2}(1^{3}D_{2})$ can be related with $M_{K_{2}(1773)}$ and $M_{K_{2}(1820)}$ by the following relation,

$$
\left(\begin{array}{cc}
\cos\theta_{K}  & \sin\theta_{K} \\
-\sin\theta_{K} & \cos\theta_{K}
\end{array}\right)
\left(\begin{array}{cc}
M^2_{K_{2}(1^{1}D_{2})} & \xi\\
\xi & M^2_{K_{2}(1^{1}D_{2})}
\end{array}\right)
\left(\begin{array}{cc}
\cos\theta_{K}  & -\sin\theta_{K} \\
\sin\theta_{K} & \cos\theta_{K}
\end{array}\right)
$$
\begin{equation}
\label{Eq8}
=\left(\begin{array}{cc}
M^2_{K_{2}(1773)} & 0\\
0 & M^2_{K_{2}(1820)}
\end{array}\right),
\end{equation}
where $\xi$ represents a parameter describing the $K_{2}(1^{1}D_{2})$ and $K_{2}(1^{3}D_{2})$ mixing.
From the relations (\ref{Eq6}), (\ref{Eq7}), and (\ref{Eq8}), the mixing angle $\theta_{K}$ and the mass relations between $M_{K_{2}(1^{1}D_{2})}$ and $M_{K_{2}(1^{3}D_{2})}$ are obtained, i.e.,
\begin{equation}
\label{Eq9}
M_{K_{2}\left({1}^{1}D_{2}\right)}^{2}=M_{K_{2}(1773)}^{2} \cos ^{2} \theta_{K}+M_{K_{2}(1820)}^{2} \sin ^{2} \theta_{K},
\end{equation}
\begin{equation}
\label{Eq10}
M_{K_{2}\left({1}^{3}D_{2}\right)}^{2}=M_{K_{2}(1773)}^{2} \sin ^{2} \theta_{K}+M_{K_{2}(1820)}^{2} \cos ^{2} \theta_{K},
\end{equation}

\begin{equation}
\label{Eq11}
\cos \left(2 \theta_{K}\right)=\frac{M_{K_{2}\left({1}^{1} D_{2}\right)}^{2}-M_{K_{2}\left({1 }^{3} D_{2}\right)}^{2}}{M_{K_{2}(1773)}^{2}-M_{K_{2}(1820)}^{2}},
\end{equation}
\begin{equation}
\label{Eq12}
M_{K_{2}\left({1}^{1}D_{2}\right)}^{2}+M_{K_{2}\left({1}^{3}D_{2}\right)}^{2}=M_{K_{2}(1773)}^{2}+M_{K_{2}(1820)}^{2}.
\end{equation}

As described in the previous section, based on the mass mixing matrix of isoscalar states and the mixing of the spin-singlet and spin-triplet, we build a bridge that connects the masses of meson nonet members. In the subsequent analysis, we will determine the masses of the unknown meson states by applying relations (\ref{Eq4}), (\ref{Eq5}), and (\ref{Eq12}) to the known meson masses.

\section{Regge phenomenology and mass relations}

In this section, we use Regge theory for the assignment of meson multiplets. The Regge theory, which connects the high energy behavior of the scattering amplitude with singularities in the complex angular momentum plane of the partial wave amplitudes, was developed in the 1960s \cite{Regge:1959mz}. Recently, Regge theory has attracted renewed attention in that it can be used to predict the meson masses or to determine the quantum numbers of newly observed states in experiments \cite{Li:2004gu,Burakovsky:1998zk,Li:2007px,Guo:2008he,Anisovich:2000kxa,Masjuan:2012gc}. Regge theory indicates that mesons are associated with Regge poles, which move in the complex angular momentum plane as a function of energy. The plots of Regge trajectories of hadrons in the $(J, M^{2})$ plane are usually called Chew-Frautschi plots, where $J$ and $M$ are the total spins and the masses of the hadrons, respectively. According to the Chew-Frautschi conjecture, the poles fall onto linear trajectories in the $(J, M^{2})$ plane,

\begin{equation}
\label{Eq13}
J=\alpha_{n\bar{n}N}(0)+\alpha'_{n\bar{n}N}M^2_{n\bar{n}N},
\end{equation}
\begin{equation}
\label{Eq14}
J=\alpha_{n\bar{s}N}(0)+\alpha'_{n\bar{s}N}M^2_{n\bar{s}N},
\end{equation}
\begin{equation}
\label{Eq15}
J=\alpha_{s\bar{s}N}(0)+\alpha'_{s\bar{s}N}M^2_{s\bar{s}N},
\end{equation}
where $N$ is the radial quantum number. The $\alpha'$ and $\alpha$ are the slope and intercept of the Regge trajectory, respectively. In this work, the intercept and slope can be expressed as
\begin{equation}
\label{Eq16}
\alpha_{n\bar{n}N}(0)+\alpha_{s\bar{s}N}(0)=2\alpha_{n\bar{s}N}(0),
\end{equation}
\begin{equation}
\label{Eq17}
\frac{1}{\alpha'_{n\bar{n}N}}+\frac{1}{\alpha'_{s\bar{s}N}}=\frac{2}{\alpha'_{n\bar{s}N}}.
\end{equation}

The intercept relation was derived from the dual-resonance model \cite{Berezinsky:1969erk}, and is satisfied in two-dimensional QCD \cite{Brower:1977as}, the dual-analytic model \cite{Kobylinsky:1978db}, the quark bremsstrahlung model \cite{Dixit:1979mz}. The slope relation (\ref{Eq17}) was obtained in the framework of topological expansion and the $q\bar{q}$-string picture of hadrons \cite{Kaidalov:1980bq}.

From relations (\ref{Eq13})-(\ref{Eq17}), one has
\begin{equation}
\label{Eq18}
M^{2}_{n\bar{n}N}\alpha'_{n\bar{n}N}+M^{2}_{s\bar{s}N}\alpha'_{s\bar{s}N}=M^{2}_{n\bar{s}N}\alpha'_{n\bar{s}N}.
\end{equation}

Based on the assumption that the slopes of parity partner trajectories coincide and are independent of charge conjugation \cite{Li:2004gu,Brisudova:1999ut}, that is to say,
\begin{equation}
  \left\{
   \begin{array}{c}
\alpha'_{n\bar{n}}(1^{1}P_{1})=\alpha'_{n\bar{n}}(1^{1}D_{2})=\alpha'_{n\bar{n}}(1^{3}D_{2}) \\ \\
\alpha'_{n\bar{s}}(1^{1}P_{1})=\alpha'_{n\bar{s}}(1^{1}D_{2})=\alpha'_{n\bar{s}}(1^{3}D_{2}) \\ \\
\alpha'_{s\bar{s}}(1^{1}P_{1})=\alpha'_{s\bar{s}}(1^{1}D_{2})=\alpha'_{s\bar{s}}(1^{3}D_{2}) \\
    \end{array}
  \right. .
\end{equation}
we have the following relations, which are related to meson multiplets with the same regge slopes by eliminating the slopes,

$$
\frac  { 4M^{2}_{n\bar{s}({1^{1}D_{2}})}  M^{2}_{n\bar{n}({1^{1}P_{1}})}-4M^{2}_{n\bar{n}({1^{1}D_{2}})}   M^{2}_{n\bar{s}({1^{1}P_{1}})}  }
       { M^{2}_{n\bar{n}({1^{1}D_{2}})}   M^{2}_{s\bar{s}({1^{1}P_{1}})}- M^{2}_{s\bar{s}({1^{1}D_{2}})}   M^{2}_{n\bar{n}({1^{1}P_{1}})}  }
$$
\begin{equation}
\label{Eq19}
= \frac { M^{2}_{n\bar{s}({1^{1}D_{2}})} ( M^{2}_{n\bar{n}({1^{1}P_{1}})} - M^{2}_{s\bar{s}({1^{1}P_{1}})} )  -
          M^{2}_{n\bar{s}({1^{1}P_{1}})} ( M^{2}_{n\bar{n}({1^{1}D_{2}})} - M^{2}_{s\bar{s}({1^{1}D_{2}})} )         }
        { M^{2}_{n\bar{s}({1^{1}D_{2}})}   M^{2}_{s\bar{s}({1^{1}P_{1}})}- M^{2}_{s\bar{s}({1^{1}D_{2}})}   M^{2}_{n\bar{s}({1^{1}P_{1}})}   },
\end{equation}

$$
\frac  { 4M^{2}_{n\bar{s}({1^{3}D_{2}})}  M^{2}_{n\bar{n}({1^{1}P_{1}})}-4M^{2}_{n\bar{n}({1^{3}D_{2}})}   M^{2}_{n\bar{s}({1^{1}P_{1}})}  }
       { M^{2}_{n\bar{n}({1^{3}D_{2}})}   M^{2}_{s\bar{s}({1^{1}P_{1}})}- M^{2}_{s\bar{s}({1^{3}D_{2}})}   M^{2}_{n\bar{n}({1^{1}P_{1}})}  }
$$
\begin{equation}
\label{Eq20}
= \frac { M^{2}_{n\bar{s}({1^{3}D_{2}})} ( M^{2}_{n\bar{n}({1^{1}P_{1}})} - M^{2}_{s\bar{s}({1^{1}P_{1}})} )  -
          M^{2}_{n\bar{s}({1^{1}P_{1}})} ( M^{2}_{n\bar{n}({1^{3}D_{2}})} - M^{2}_{s\bar{s}({1^{3}D_{2}})} )         }
        { M^{2}_{n\bar{s}({1^{3}D_{2}})}   M^{2}_{s\bar{s}({1^{1}P_{1}})}- M^{2}_{s\bar{s}({1^{3}D_{2}})}   M^{2}_{n\bar{s}({1^{1}P_{1}})}   }.
\end{equation}

Inserting the masses of states into relations (\ref{Eq12}), (\ref{Eq19}), (\ref{Eq20}), with the aid of $M^{2}_{s\bar{s}({1^{3}D_{2}})}=2M^{2}_{n\bar{s}({1^{3}D_{2}})}-M^{2}_{n\bar{n}({1^{3}D_{2}})}$ \cite{Li:2008et,Okubo:1961jc}, we obtain the values of $M_{n\bar{s}({1^{1}D_{2}})}$, $M_{n\bar{s}({1^{3}D_{2}})}$, $M_{n\bar{n}({1^{3}D_{2}})}$, $M_{s\bar{s}({1^{3}D_{2}})}$ and list the results in Table 3 and 4. In this work, we take

\begin{equation}
\label{Eq21}
M^{2}_{n\bar{s}({1^{1}P_{1}})}=\frac{1}{2} M^{2}_{K_{1}(1270)} +\frac{1}{2} M^{2}_{K_{1}(1400)},
\end{equation}
the $K_{1A}$ and $K_{1B} $ are nearly equal $45^{\circ}$ mixtures of the $K_{1}(1270)$ and $K_{1}(1400)$ \cite{Divotgey:2013jba}.

Apart from the ground meson, the radial excitations can be estimated in the framework of Regge phenomenology.
Based on the assumption that the ground and the radial excitation have the same slopes \cite{Anisovich:2000kxa}, we obtain the following from the relations (\ref{Eq13}), (\ref{Eq14}), and (\ref{Eq15}),
\begin{equation}
\label{Eq22}
M^{2}_{n\overline{n}N}\alpha'_{n\overline{n}1}-M^{2}_{n\bar{n}1}\alpha'_{n\overline{n}1}=\alpha_{n\bar{n}1}(0)-\alpha_{n\bar{n}N}(0),
\end{equation}
\begin{equation}
\label{Eq23}
M^{2}_{s\overline{s}N}\alpha'_{s\overline{s}1}-M^{2}_{s\bar{s}1}\alpha'_{s\overline{s}1}=\alpha_{s\bar{s}1}(0)-\alpha_{s\bar{s}N}(0),
\end{equation}

\begin{equation}
\label{Eq24}
M^{2}_{n\overline{s}N}\alpha'_{n\overline{s}1}-M^{2}_{n\bar{s}1}\alpha'_{n\overline{s}1}=\alpha_{n\bar{s}1}(0)-\alpha_{n\bar{s}N}(0).
\end{equation}

In Refs. \cite{Filipponi:1997hb,Filipponi:1997vf},  Filipponi et al. indicate that the values of $\alpha_{n\bar{n}1}(0)-\alpha_{n\bar{n}N}(0)$, $\alpha_{n\bar{s}1}(0)-\alpha_{n\bar{s}N}(0)$, and $\alpha_{s\bar{s}1}(0)-\alpha_{s\bar{s}N}(0)$ depend on the constituent quark masses through the combination $m_{i} + m_{j}$ ($m_{i}$ and $m_{j}$ are the constituent masses of quark and antiquark). In this case, a factor $f_{i\overline{j}}(m_{i}+m_{j})$ is introduced into relations (\ref{Eq22})-(\ref{Eq24}) \cite{Li:2007px,Liu:2010zzd}. Then these relations are expressed as

\begin{equation}
\label{Eq25}
M^{2}_{n\bar{n}N}=M^{2}_{n\bar{n}1}+\frac{(N-1)}{\alpha'_{n\bar{n}}}(1+f_{n\bar{n}}(m_{n}+m_{n})),
\end{equation}

\begin{equation}
\label{Eq26}
M^{2}_{n\bar{s}N}=M^{2}_{n\bar{s}1}+\frac{(N-1)}{\alpha'_{n\bar{s}}}(1+f_{n\bar{s}}(m_{n}+m_{s})),
\end{equation}

\begin{equation}
\label{Eq27}
M^{2}_{s\bar{s}N}=M^{2}_{s\bar{s}1}+\frac{(N-1)}{\alpha'_{s\bar{s}}}(1+f_{s\bar{s}}(m_{s}+m_{s})).
\end{equation}

In this work, the parameters used as input are taken from our previous work (Table 2) \cite{Liu:2010zzd}.

\begin{center}
\indent\\ \footnotesize Table 2. The slopes (GeV$^{-2}$) and the parameters $f_{nn}$ ,$f_{ns}$ and $f_{ss}$ (GeV$^{-1}$) of relations (19), (20), and (21).
\label{Tab:t2}
\begin{tabular}{lllllll}
\hline
\hline
parameters     &$\alpha'_{n\bar{n}}$ & $\alpha'_{n\bar{s}}$ & $\alpha'_{s\bar{s}}$  &$f_{nn}$ & $f_{ns}$ & $f_{ss}$
\\
\hline
value         &  $0.7218$   &$0.6613$    & $0.76902$     & $0.3556$ & $0.1376$  & $0.2219$
\\
\hline

\end{tabular}

\end{center}

The results are shown in Table 3, 4 and Figure 1, 2.

\begin{center}
\indent\\ \footnotesize Table 3. The radial excitation masses of the $1^{1}D_{2}$ multiplet in this work are compared with the other predictions (in units of MeV). This work (1) and (2) are the solutions of relation (20). The mass used as input for our calculation is shown in boldface.
\label{Tab:t3}
\begin{tabular}{llllllllllll}
\hline
\hline
 $1^{1}D_{2}$         &$N$  &This work(1) & This work(2) &\cite{Li:2020xzs} &\cite{Xiao:2019qhl} &\cite{Ebert:2009ub}&\cite{Pang:2017dlw}& \cite{Godfrey:1985xj}
\\
\hline
                &1  & \boldmath{$1670.6^{+2.9}_{-1.2}$} &                  &     &     &1643 &    &1680
\\
$M_{n\bar{n}}$  &2  &$2111.9{\pm2.3}$               &                  &     &     &1960 &    &2130
\\
                &3  &$2476.1{\pm2.0}$             &                  &     &     &2216 &    &
\\
\hline
                &1  &$1761.7\pm 5.6$                    & $1767.4\pm 5.9$  &     &     &1709 &1778&1780
\\
$M_{n\bar{s}}$  &2  &$2188.7\pm 4.6$                    & $2193.3\pm 4.8$  &     &     &2066 &2121&2230
\\
                &3 &$2545.0\pm 3.8$                    & $2549.0\pm 4.1$  &     &     &     &2380&
\\
\hline
                &1  &$1849.0\pm 8.1$                    &                  &1825 &1893 &1909 &    &1890
\\
$M_{s\bar{s}}$  &2  &$2262.4\pm 6.7$                    &                  &2282 &2336 &2321 &    &
\\
                &3  &$2611.1\pm 5.8$                    &                  &2685 &2723 &2662 &    &
\\
\hline
\hline
\end{tabular}
\end{center}

\begin{center}
\indent\\ \footnotesize Table 4. The radial excitation masses of the $1^{3}D_{2}$ multiplet in this work are compared with the other predictions (in units of MeV). This work (1) and (2) are the solutions of relation (21).
\label{Tab:t4}
\begin{tabular}{llllllllllll}
\hline
\hline
 $1^{3}D_{2}$          &$N$      &This work(1) & This work(2) &\cite{Li:2020xzs} &\cite{Xiao:2019qhl} &\cite{Ebert:2009ub}&\cite{Pang:2017dlw}& \cite{Godfrey:1985xj}
\\
\hline
                &1 &$1687.7\pm 15.5$ & $1692.3\pm 15.6$  &     &     &1661 &    &1700
\\
$M_{n\bar{n}}$  &2  &$2125.9\pm 12.3$ & $2129.6\pm 12.4$  &     &     &1983 &    &2150
\\
                &3  &$2488.1\pm 10.6$ & $2491.2\pm 10.6$  &     &     &2241 &    &
\\
\hline
                &1  &$1824.8\pm 9.6$ & $1829.7\pm 9.8$    &     &     &1824 &1789&1810
\\
$M_{n\bar{s}}$  &2  &$2236.8\pm 7.9$ & $2243.8\pm 8.0$    &     &     &2163 &2131&2260
\\
                &3  &$2589.1\pm 6.8$ & $2592.6\pm 7.0$    &     &     &     &2388&
\\
\hline
                &1  &$1952.4\pm 13.5$ & $1957.5\pm 13.6$  &1840 &1904 &     &    &1910
\\
$M_{s\bar{s}}$  &2  &$2347.7\pm 11.3$ & $2351.9\pm 11.4$  &2297 &2348 &     &    &
\\
                &3 &$2685.3\pm 9.9$ & $2689.0\pm 10.1$   &2701 &2734 &     &    &
\\
\hline
\hline
\end{tabular}
\end{center}

\begin{figure}[htb]
\centerline{
\subfigure {
\includegraphics[width=0.6\columnwidth]{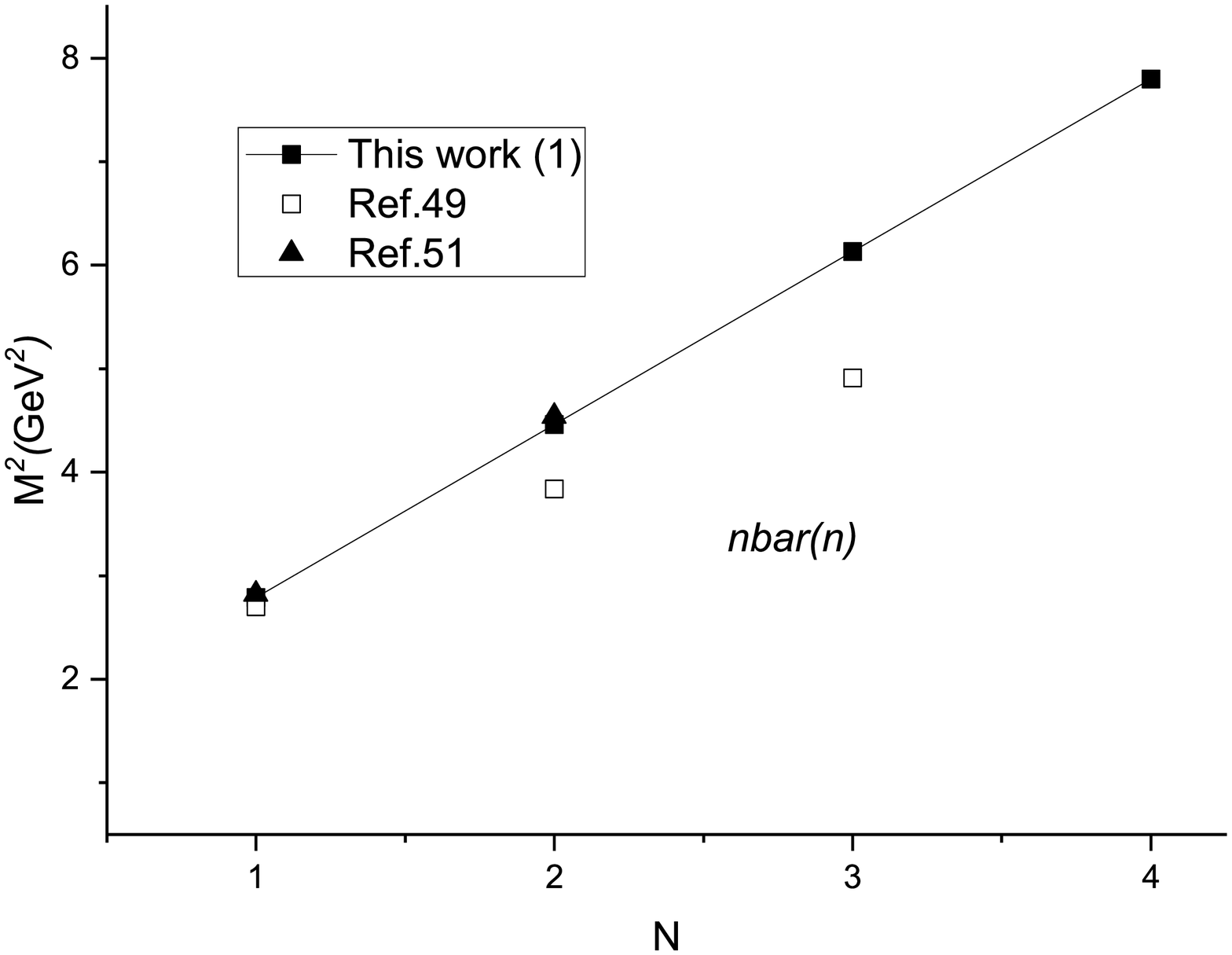}
}
}
\centerline{
\subfigure {
\includegraphics[width=0.6\columnwidth]{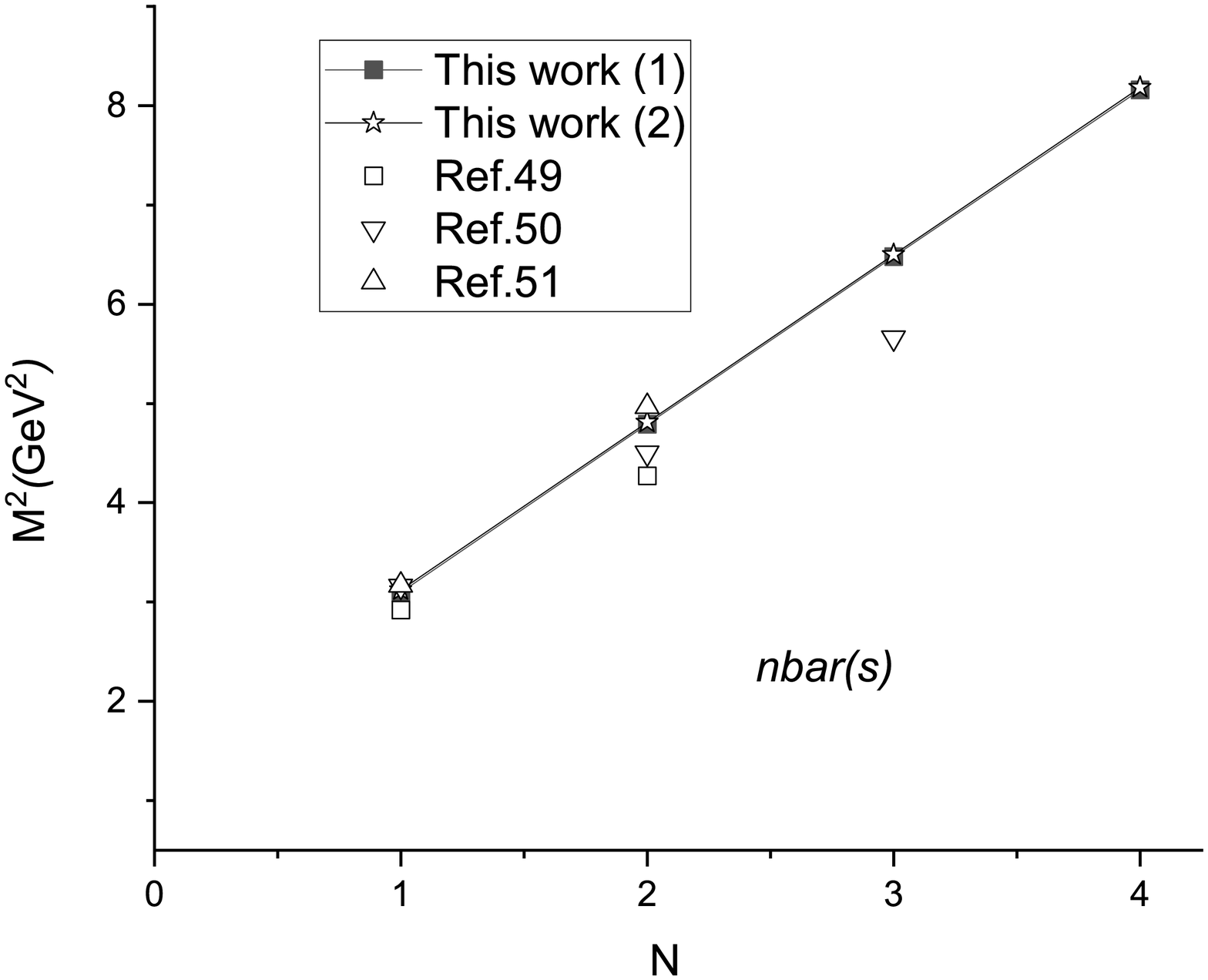}
}
}
\centerline{
\subfigure {
\includegraphics[width=0.6\columnwidth]{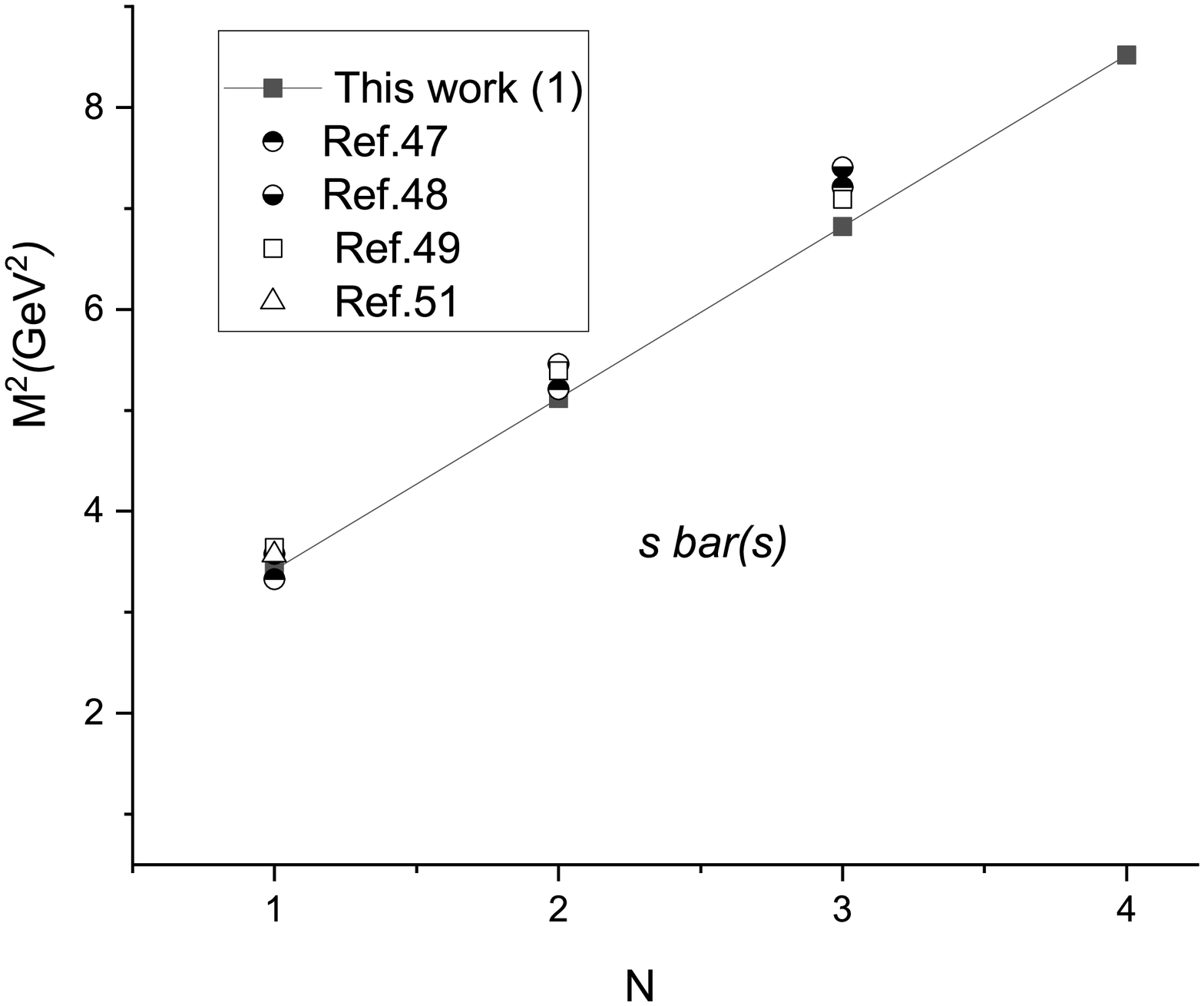}
}
}
\caption{ The radial excitation of $N^{1}D_{2}$ meson nonet. }
\label{Fig:F1}
\end{figure}

\begin{figure}[htb]
\centerline{
\subfigure {
\includegraphics[width=0.6\columnwidth]{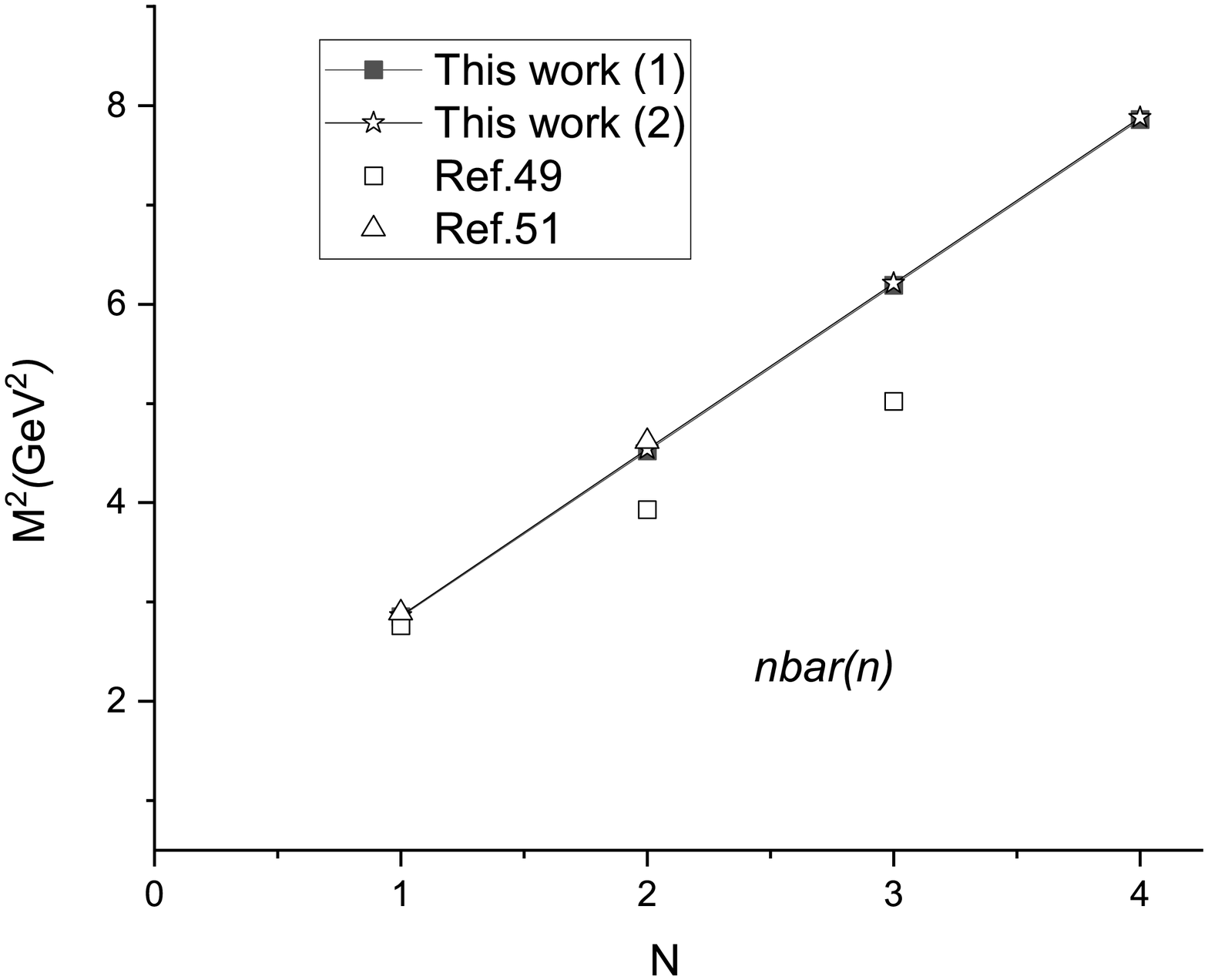}
}}
\centerline{
\subfigure {
\includegraphics[width=0.6\columnwidth]{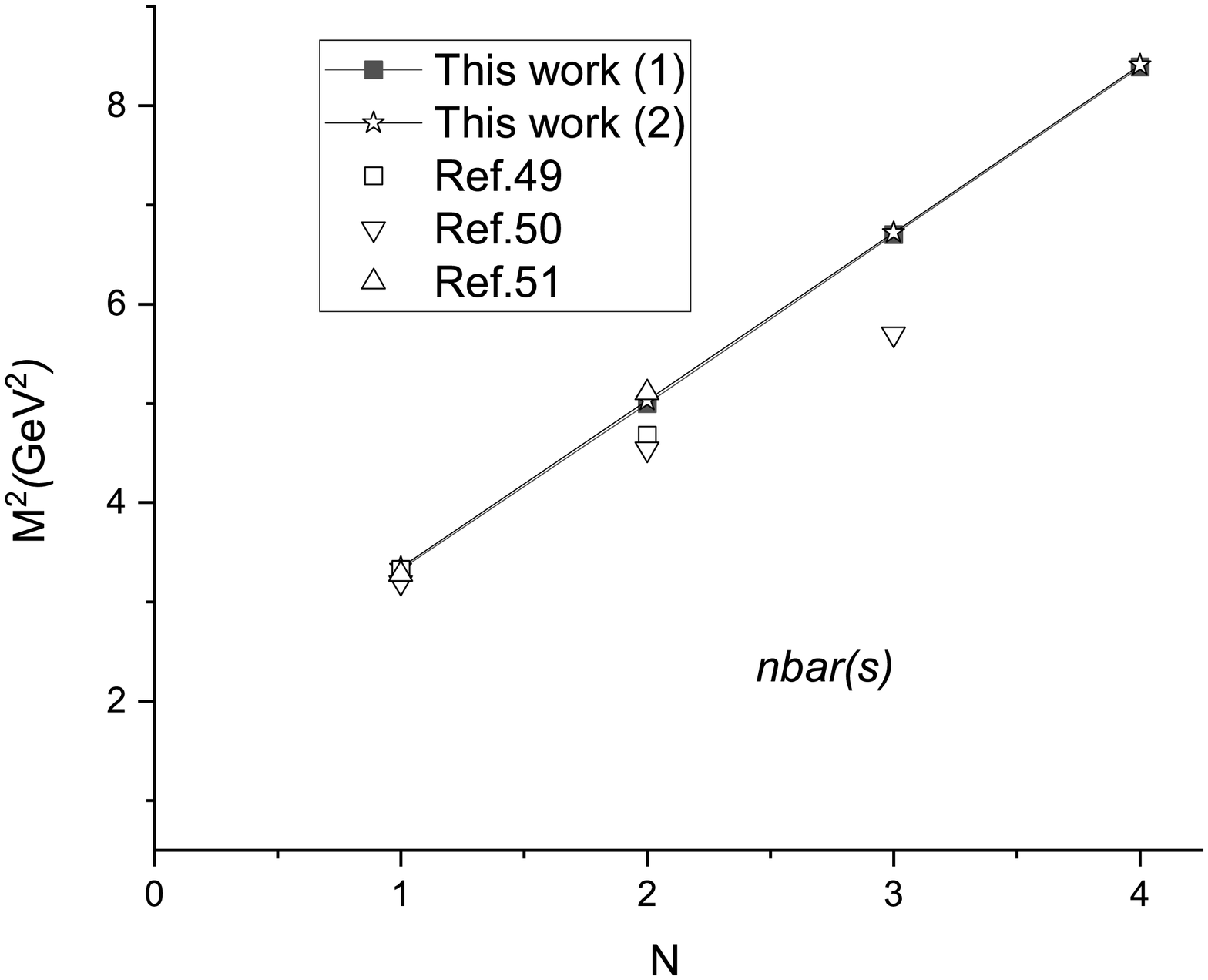}
}}
\centerline{
\subfigure {
\includegraphics[width=0.6\columnwidth]{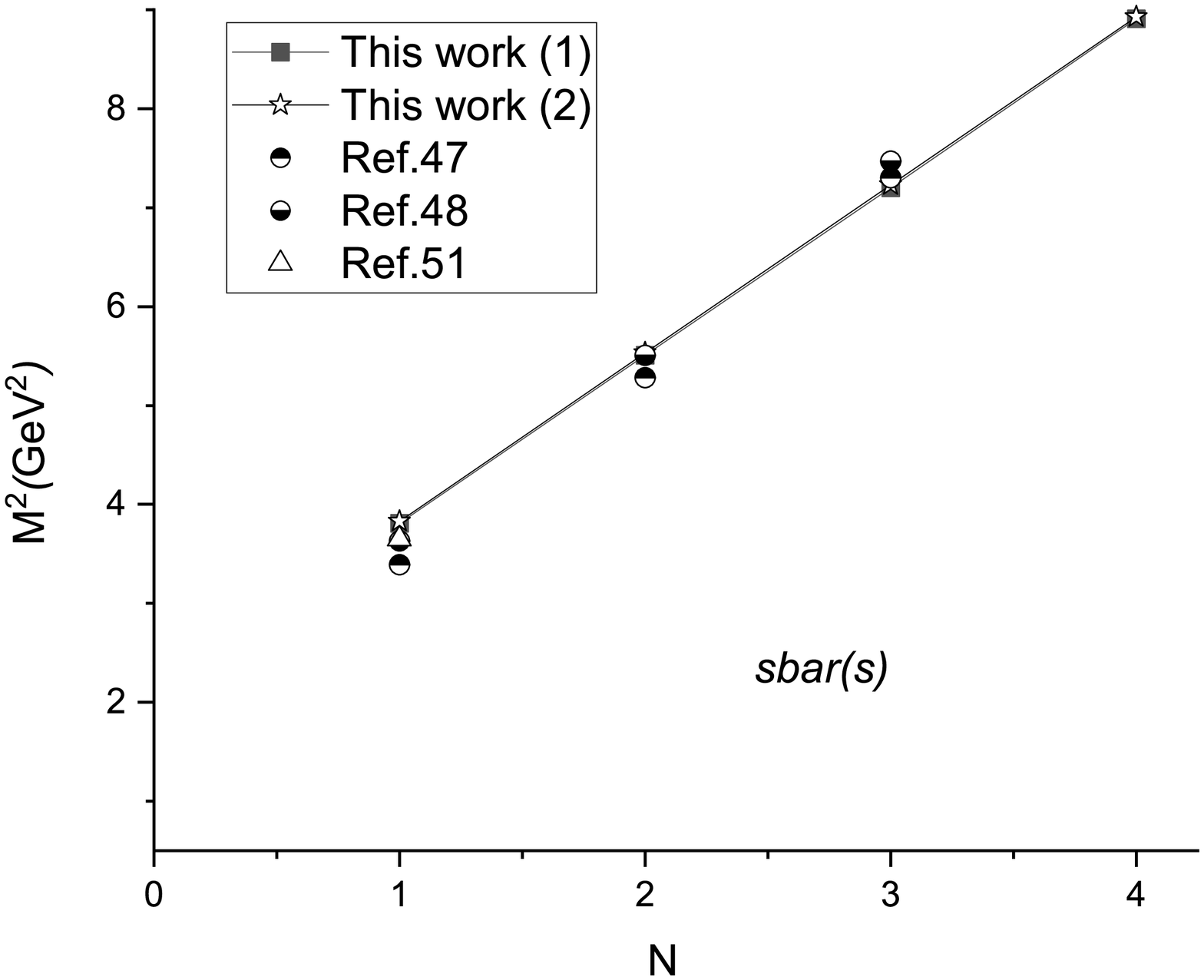}
}}
\caption{ The radial excitation of $N^{1}D_{2}$ meson nonet. }
\label{Fig:F2}
\end{figure}

\section{Conclusion}

In this work, we investigate the mass spectrum of $1^{1}D_{2}$ and $1^{3}D_{2}$ meson nonets in the framework of the meson mass matrix and Regge phenomenology.
The mass of $1^{1}D_{2}(s\bar{s})$ is determined to be $1849.0$ MeV, which agrees with the state $\eta_{2}(1870)$ with a mass of $1842\pm 8$ MeV in PDG.
The consistency of masses also indicates that $\eta_{2}(1870)$ is the candidate for $1^{1}D_{2}$ meson nonet with a small mixing with $n\bar{n}$ component, the corresponding results need to be further confirmed in the experiment. Apart from the ground meson, the masses of radial excitations of $1^{1}D_{2}$ are obtained. There are five isovector states with $J^{PC}=2^{-+}$ ($\pi_{2}(1670)$, $\pi_{2}(1880)$, $\pi_{2}(2050)$, $\pi_{2}(2100)$, $\pi_{2}(2285)$) are listed in PDG. The $\pi_{2}(2100)$ is omitted from the summary table, the states $\pi_{2}(2050)$ and $\pi_{2}(2285)$ are classified as ``further states''. In this work, the isovector of $2^{1}D_{2}$ mass is determined to be $2111.9$ MeV, which is consistent with the mass of $\pi_{2}(2100)$ ($2090\pm29$ MeV). We suggest that $\pi_{2}(2100)$ would be the plausible candidate of $2^{1}D_{2}$ meson state.
For the $1^{3}D_{2}$ meson nonet, only the kaon $K_{2}(1820)$ is found, and it will mix with $K_{2}(1770)$ state. In PDG, the four states $\rho_{2}(1940)$, $\omega_{2}(1975)$, $\omega_{2}(2195)$, $\rho_{2}(2225)$ with $J^{PC}=2^{--}$ are interpreted as ``further states''. In this work, we find that there is still a large deviation between our results and the masses of these states. Till now, since the experimental information of $1^{3}D_{2}$ meson nonet is not enough, it is necessary to conduct a systematic and complete study for these states. The results may be useful for the discovery of unobserved states in the experiments (BESIII, CMD-3, and COMPASS).

\end{document}